\documentclass[twocolumn,showpacs,amsmath,amssymb,prl]{revtex4}
\usepackage{amsxtra}
\usepackage{amssymb}
\usepackage{amsmath}
\usepackage{graphicx}

\newcommand {\rhovec}{\ensuremath \boldsymbol{\rho}}

\begin{document}
\title{Computational Ghost Imaging}
\date{\today}
\author{Jeffrey H. Shapiro}
\affiliation{Massachusetts Institute of Technology, Research Laboratory of Electronics, Cambridge, Massachusetts 02139, USA}

\begin{abstract}
Ghost-imaging experiments correlate the outputs from two photodetectors:  a high spatial-resolution (scanning pinhole or CCD camera) detector that measures a field which has not interacted with the object to be imaged, and a bucket (single-pixel) detector that collects a field that has interacted with the object.  We describe a computational ghost-imaging arrangement that uses only a single-pixel detector.  This configuration affords background-free imagery in the narrowband limit and a 3D sectioning capability.  It clearly indicates the classical nature of ghost-image formation.  
\end{abstract}
\pacs{42.30.Va, 42.50.Ar, 42.25.Kb}
\maketitle 
Ghost imaging is the acquisition of object information by means of photocurrent correlation measurements. Its first demonstration utilized a biphoton source, thus the image was interpreted as a quantum phenomenon owing to the entanglement of the source photons \cite{Pittman}.  Experimental \cite{Valencia,Ferri} and theoretical \cite{Gatti:three,Gatti} work later demonstrated that ghost imaging could be performed with pseudothermal light. Whereas the biphoton requires a quantum description for its photodetection statistics, pseudothermal light can be regarded as a classical electromagnetic wave whose photodetection statistics can be treated via the semiclassical (shot-noise) theory of photodetection \cite{Mandel}.  This disparity has sparked interest in the physics of ghost imaging \cite{DAngelo,Cai,Cai2,Scarcelli}.  Recently \cite{ErkmenShapiro}, we established a Gaussian-state analysis of ghost imaging that unified prior work on biphoton and pseudothermal sources.  Our analysis indicated that ghost-image formation is intrinsically due to classical coherence propagation, with the principal advantage afforded by the biphoton state being  high-contrast imagery in the wideband limit. Other recent work \cite{Meyers,Shih}, however, has claimed that pseudothermal-light ghost imaging is a fundamentally quantum phenomenon---one of nonlocal two-photon interference---that \em cannot\/\rm\ be explained in terms of intensity-fluctuation correlations.    In this Letter we shall show that ghost imaging can be accomplished with only \em one\/\rm\ detector, viz., the bucket detector that collects a single pixel of light which has been transmitted through or reflected from the object.  As only one light beam and one photodetector are required, this imaging configuration \em cannot\/\rm\ depend on nonlocal two-photon interference.  Moreover, it affords background-free imagery in the narrowband limit and a 3D sectioning capability.

We begin with a review of pseudothermal-light lensless ghost imaging, based on \cite{ErkmenShapiro},  using classical electromagnetism and semiclassical photodetection theory.  Consider the setup shown in Fig.~1. Here, $E_{S}(\rhovec,t)e^{-i \omega_{0}t}$ and $E_{R}(\rhovec,t)e^{-i \omega_{0}t}$ are scalar, positive frequency, classical signal ($S$) and reference ($R$) fields that are normalized to photon-units and have center frequency $\omega_0$.  They are $z$-propagating with $\rhovec$ being the transverse coordinate with respect to their optical axes.  More importantly, they are the outputs from  50/50 beam splitting of $E(\rhovec,t)$, a continuous-wave (cw) laser beam that has been transmitted through a rotating ground-glass diffuser.  The signal and reference undergo quasimonochromatic paraxial diffraction over $L$-m-long free-space paths, yielding measurement-plane fields \cite{Goodman}
\begin{equation}
E_{\ell}(\rhovec,t) \!=\! \int\! d \rhovec'\, E_{m}\big(\rhovec',t\!-\!L/c \big) \frac{k_{0} e^{i k_{0} (L+ |\rhovec-\rhovec'|^2/2L)}}{i 2 \pi L}, \label{FS:prop}
\end{equation} 
where $(\ell,m) = (1,S)$ or $(2,R)$,  $c$ is the speed of light, and $k_{0} = \omega_{0}/c = 2\pi/\lambda_0$. The field $E_{1}(\rhovec,t)$ illuminates a shot-noise limited pinhole photodetector centered at $\rhovec_{1}$ with sensitive region $\rhovec \in {\cal{A}}_1$. The field $E_{2}(\rhovec,t)$ illuminates an amplitude-transmission mask $T(\rhovec)$, located immediately in front of a shot-noise limited bucket photodetector with sensitive region $\rhovec \in \mathcal{A}_{2}$. The product of the photocurrents from these detectors is time averaged to estimate their ensemble-average cross correlation, $C(\rhovec_{1})$.  This process is repeated, as $\rhovec_{1}$ is scanned over the plane, to obtain the ghost image of the object's intensity transmission $|T(\rhovec)|^2$ \cite{footnote1}.
 \begin{figure}[t]
\begin{center}
\includegraphics[width= 3.25in]{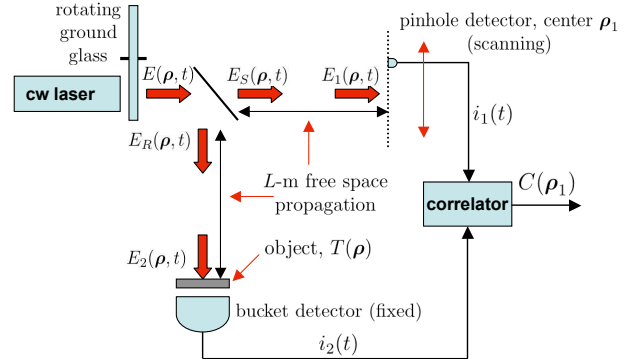}
\end{center}
\vspace*{-.25in}
\caption{(Color online) Pseudothermal ghost-imaging setup.} \label{GI:propagation}
\end{figure}
We then have that
\begin{eqnarray}
C(\rhovec_1) &=&  \left\langle \int\!d\tau_1\, q\eta_1 P_1(t-\tau_1)h_1(\tau_1)\right. \nonumber\\
&\times& \left. \int\!d\tau_2\, q\eta_2 P_2(t-\tau_2)h_2(\tau_2) \right\rangle.
\end{eqnarray}
In this expression:  $q$ is the electron charge; $\eta_\ell$ is the quantum efficiency of the pinhole ($\ell = 1$) and bucket ($\ell =2$) detectors; 
$P_1(t) = \int_{\mathcal{A}_1}\!d\rhovec\,|E_1(\rhovec,t)|^2$ and $P_2(t) = \int_{\mathcal{A}_2}\!d\rhovec\,|E_2(\rhovec,t)|^2|T(\rhovec)|^2$ are the photon fluxes impinging on detectors 1 and 2; and $h_\ell(t)$ is a baseband impulse response representing the finite response time of the photodetector $\ell$.  We shall assume that the pinhole is small enough that $P_1(t) \approx A_1|E(\rhovec_1,t)|^2$, where $A_1$ is that detector's photosensitive area.  In terms of the intensities (photon-flux densities) 
$I_1(\rhovec,t)  = |E_1(\rhovec,t)|^2$,  and $I_2(\rhovec,t) = 
|E_2(\rhovec,t)|^2|T(\rhovec)|^2$
that illuminate the photodetectors we then have that
\begin{eqnarray}
\lefteqn{\hspace*{-.25in}\langle P_1(t_1)P_2(t_2)\rangle = A_1\langle I_1(\rhovec_1,t_1)\rangle\!\int_{\mathcal{A}_2}\!d\rhovec\,  \langle I_2(\rhovec,t_2)\rangle } 
\nonumber\\
&+& A_1\int_{\mathcal{A}_2}\!d\rhovec\, \langle \Delta I_1(\rhovec_1,t_1)\Delta I_2(\rhovec,t_2)\rangle,
\end{eqnarray}
where $\Delta I_\ell(\rhovec,t) \equiv I_\ell(\rhovec,t) - \langle I_\ell(\rhovec,t)\rangle$ is the intensity fluctuation.  The first term on the right gives rise to a featureless background, while the second term leads to the ghost image, as we now show for a Gaussian-Schell model of pseudothermal illumination.  

Let $E(\rhovec,t)$ be a zero-mean, cross-spectrally pure \cite{coherence_separability}, complex-valued Gaussian random process that is characterized by its phase-insensitive correlation function 
\begin{equation}
\langle E^*(\rhovec_1,t_1)E(\rhovec_2,t_2) \rangle = K(\rhovec_1, \rhovec_2) R(t_2-t_1) \label{CORR:PIS},
\end{equation}
where $R(0) = 1$.  
Using $E_m(\rhovec,t) = E(\rhovec,t)/\sqrt{2}$, for $m = S,R$, and Eq.~(\ref{FS:prop}) we have that $\{E_\ell(\rhovec,t) : \ell = 1,2\}$, is a pair of completely correlated, zero-mean, complex-valued Gaussian random processes that are characterized by the following auto- and cross-correlation functions:
\begin{eqnarray}
\langle E^*_\ell(\rhovec_1,t_1)E_\ell(\rhovec_2,t_2)\rangle &=& 
\langle E^*_1(\rhovec_1,t_1)E_2(\rhovec_2,t_2)\rangle \\[.12in]
&=& K'(\rhovec_1,\rhovec_2)R(t_2-t_1),
\label{impliedCorr}
\end{eqnarray}
for $\ell = 1,2$.  Given an explicit $K(\rhovec_1,\rhovec_2)$ for the spatial auto-correlation function of $E(\rhovec,t)$, it is a simple matter, in principle, to calculate $K'(\rhovec_1,\rhovec_2)$ via standard coherence-propagation theory \cite{Mandel}.  We then have that 
$\langle I_1(\rhovec,t) \rangle = K'(\rhovec,\rhovec)$ and $\langle I_2(\rhovec,t)\rangle = 
K'(\rhovec,\rhovec)|T(\rhovec)|^2$.   More importantly, the moment-factoring theorem for Gaussian random processes \cite{Wozencraft} implies that
\begin{eqnarray}
\langle\Delta I_1(\rhovec_1,t_1)\Delta I_2(\rhovec_2,t_2)\rangle &=&  
|K'(\rhovec_1,\rhovec_2)|^2|R(t_2-t_1)|^2\nonumber \\ &\times& |T(\rhovec_2)|^2.
\label{fluct}
\end{eqnarray}

In the far field (when $k_0a_0\rho_0/2L \ll 1$) the Gaussian-Schell model correlation function for $E(\rhovec,t)$, 
\begin{equation}
K(\rhovec_1,\rhovec_2) = \frac{2P}{\pi a_0^2}e^{-(|\rhovec_1|^2 + |\rhovec_2|^2)/a_0^2 -|\rhovec_1-\rhovec_2|^2/2\rho_0^2},
\end{equation}
with $\rho_0 \ll a_0$, yields
\begin{eqnarray}
K'(\rhovec_1,\rhovec_2) &=& \frac{P}{\pi a_L^2}e^{ik_0(|\rhovec_2|^2 - |\rhovec_1|^2)/2L}
\nonumber \\ &\times& e^{-(|\rhovec_1|^2 + |\rhovec_2|^2)/a_L^2 -|\rhovec_1-\rhovec_2|^2/2\rho_L^2},
\label{GSoutput}
\end{eqnarray}
with $a_L = 2L/k_0\rho_0$ and $\rho_L = 2L/k_0a_0$.  Physically, $a_z$ and $\rho_z$ are the intensity radii and coherence radii of the fields at $z=0$ and $z=L$, so that the preceding behavior represents the familiar situation for partially coherent light in which the far-field intensity radius is controlled by the source's coherence radius and the far-field coherence radius is controlled by the source's intensity radius.  

Suppose that the photodetector impulse responses $h_\ell(t)$ have response times that are much shorter than the field's coherence time, $T_0$, i.e., we are in the narrowband regime.  The Gaussian-Schell model source then leads to 
\begin{eqnarray}
C(\rhovec_{1}) &=& C_{0}(\rhovec_1) + q^2\eta_1\eta_2 A_1\!\left(\frac{P}{\pi a_L^2}\right)^2 e^{-2|\rhovec_1|^2/a_L^2}\nonumber \\[.08in]
&\times & \int_{\mathcal{A}_{2}}\! d\rhovec\,  e^{-2|\rhovec|^2/a_L^2-|\rhovec_1-\rhovec|^2/\rho_L^2} |T(\rhovec)|^{2},\label{GI:corr}
\end{eqnarray}
where we have assumed that $\int\!dt\,h_\ell(t) = 1$, and
\begin{eqnarray}
C_0(\rhovec_1) &\equiv& q^2\eta_1\eta_2A_1\!\left(\frac{P}{\pi a_L^2}\right)^2e^{-2|\rhovec_1|^2/a_L^2} \nonumber\\[.08in]
&\times& \int_{\mathcal{A}_2}\!d\rhovec\, e^{-2|\rhovec|^2/a_L^2}|T(\rhovec)|^2.
\end{eqnarray}
When $T(\rhovec)$ and $\rhovec_1$ are space limited to a radius much smaller than $a_L$ about the origin, $C(\rhovec_1)$ is comprised of a constant background term,
\begin{equation}
C_0 = q^2\eta_1\eta_2\!\left(\frac{P}{\pi a_L^2}\right)^2\int_{\mathcal{A}_2}\!d\rhovec\, |T(\rhovec)|^2,
\label{C0}
\end{equation}
plus the ghost-image term
\begin{equation}
C_1(\rhovec_1) \equiv q^2\eta_1\eta_2A_1\!\left(\frac{P}{\pi a_L^2}\right)^2\!\int_{\mathcal{A}_2}\!d\rhovec\,
e^{-|\rhovec_1-\rhovec|^2/\rho_L^2}|T(\rhovec)|^2.
\label{Gimage}
\end{equation}
Equations~(\ref{C0}) and (\ref{Gimage}) summarize the key elements of pseudothermal ghost imaging.  Within an object-plane region whose spatial extent is small compared to $\lambda_0L/\rho_0$, we obtain a pseudothermal-light ghost image with spatial resolution $\sim$$\lambda_0L/a_0$ \cite{MeyersRes} that is embedded in a featureless background \cite{background}.  The background term can be eliminated by employing a zero-frequency (DC) block between one or both of the photodetectors and the correlator shown in Fig.~1, as done in the experiment reported in \cite{Scarcelli}. 

With the preceding analysis in hand, it becomes a simple matter to walk our way through to a single-pixel ghost imager.  First, rather than use laser light transmitted through a rotating ground glass as the source of a narrowband, spatially-incoherent $E(\rhovec,t)$, let us employ the configuration shown in Fig.~2.  Here, we transmit a cw laser beam through a spatial light modulator (SLM) whose inputs are chosen to create the desired coherence behavior.  Specifically, we assume an idealized SLM consisting of $d\times d$ pixels arranged in a $(2M+1)\times (2M+1)$ array with 100\% fill factor within a $D\times D$ opaque pupil, where $D = (2M+1)d$ and $M \gg 1$.  We use this SLM to impose a phase $\phi_{nm}(t)$ on the light transmitted through pixel $(n.m)$, with $\{e^{i\phi_{nm}(t)} : -M \le n,m\le M\}$ being independent identically-distributed (iid) random processes obeying $\langle e^{i\phi_{nm}(t)}\rangle = 0$ and 
$\langle e^{i([\phi_{nm}(t_2)-\phi_{jk}(t_1)]} \rangle = \delta_{jn}\delta_{km}e^{-|t_2-t_1|/T_0},$
where the coherence time $T_0$ is long compared to the response times of the $h_\ell(t)$ \cite{telegraph}.
 \begin{figure}[t]
\begin{center}
\includegraphics[width= 3.25in]{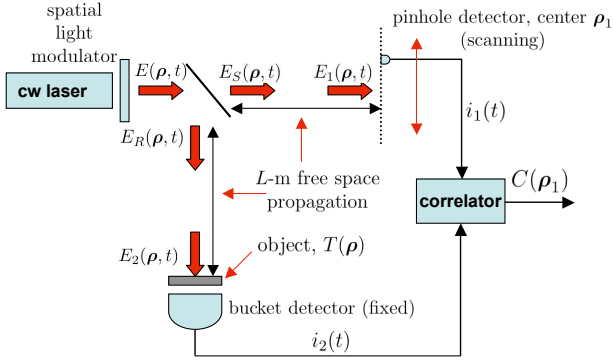}
\end{center}
\vspace*{-.25in}
\caption{(Color online) SLM ghost-imaging setup.} 
\end{figure}

In the far field, i.e., when $k_0dD/L \ll 1$, the preceding $E(\rhovec,t)$ leads to $E_1(\rhovec,t)$ and $E_2(\rhovec,t)$ that are zero-mean fields whose correlation functions satisfy Eq.~(\ref{impliedCorr}) with $R(\tau) = e^{-|\tau|/T_0}$ and 
\begin{eqnarray}
\lefteqn{K'(\rhovec_1.\rhovec_2) = \frac{P}{2}\left(\frac{d^2}{D\lambda_0L}\right)^2\! 
e^{ik_0(|\rhovec_2|^2-|\rhovec_1|^2)/2L}} 
\nonumber \\[.08in]
&\times& \left(\prod_{u = x,y}
\frac{\sin(k_0du_1/2L)}{k_0du_1/2L}\frac{\sin(k_0du_2/2L)}{k_0du_2/2L}\right) \nonumber \\[.08in]
&\times& \left(\prod_{u = x,y}
\frac{\sin[k_0D(u_1-u_2)/2L]}{\sin[k_0d(u_1-u_2)/2L]}\right).
\end{eqnarray} 
Although it is not a Gaussian-Schell form, the preceding spatial correlation function has an intensity width $\sim$$\lambda_0L/d$ and a coherence length $\sim$$\lambda_0L/D$, behavior which is qualitatively similar to what we saw earlier if we identify $d\approx \rho_0$ and $D \approx a_0$.  Furthermore, Central Limit Theorem considerations imply that $E_1(\rhovec,t)$ and $E_2(\rhovec,t)$ may be taken to be jointly Gaussian.  Hence our Fig.~2 configuration will produce a ghost image of spatial resolution $\lambda_0L/D$ within a spatial region of width $\lambda_0L/d$ embedded in a background by virtue of 
\begin{eqnarray}
C(\rhovec_1) &=& q^2\eta_1\eta_2A_1K'(\rhovec_1,\rhovec_1)\int_{\mathcal{A}_2}\!d\rhovec\,
K'(\rhovec,\rhovec)|T(\rhovec)|^2 \nonumber \\
&+&
q^2\eta_1\eta_2A_1\int_{\mathcal{A}_2}\!d\rhovec\,|K'(\rhovec_1,\rhovec)|^2|T(\rhovec)|^2.
\end{eqnarray} 
As before, the background term can be suppressed, if desired, by means of a DC block.  

To realize the Fig.~2 ghost imager we could use noise generators to drive the SLM in a way that approximates the preceding statistics.  It is more interesting, for what will follow, to suppose that deterministic driving functions are employed to achieve the same objective.  What we want at the SLM's output is a narrowband, zero-mean field whose spatial coherence---inferred now from a time average, rather than an ensemble average, because there is no true randomness---is limited to field points within a single pixel.  Sinusoidal modulation, $\phi_{nm}(t) = \Phi\cos(\Omega_0 + \Delta\Omega_{nm}t)$, with different $\Delta\Omega_{nm}$ for each pixel will work.  Let $\langle \cdot \rangle_{T_a}$ denote time averaging over the $T_a$-sec interval employed in obtaining the ghost image.  We have that
$\langle e^{i\phi_{nm}(t)}\rangle_{T_a} \approx J_0(\Phi) \approx 0$, where $J_0$ is the zeroth-order Bessel function of the first kind, when $(\Omega_0 + \Delta\Omega_{nm})T_a \gg 2\pi$ and $\Phi \gg \pi$.  With the additional condition $|\Delta\Omega_{nm}| \ll \Omega_0$ we have
$\langle e^{i[\phi_{nm}(t) - \phi_{jk}(t)]}\rangle_{T_a} \approx J_0(2\Phi) \approx 0$, unless $j=n$ and $k=m$.
Furthermore, the output field will satisfy our narrowband requirement if the modulation periods $2\pi/(\Omega_0 + \Delta\Omega_{nm})$ are all much longer that the response times of the $h_\ell(t)$.  Thus, this deterministically-modulated source will also yield a ghost image of spatial resolution $\lambda_0L/D$ within a spatial region of width $\lambda_0L/d$ embedded in a background that can be suppressed by means of a DC block.  

At this point, the notion of computational ghost imaging---in which we only use the bucket detector---is easily understood, see Fig.~3.  We use deterministic modulation of a cw laser beam to create the field $E_2(\rhovec,t)$ that illuminates the object transparency, and, as usual, we collect the light that is transmitted through the transparency with a bucket (single-pixel) detector \cite{reflect}.  Knowing the deterministic modulation applied to the original cw laser beam allows us to use diffraction theory to \em compute\/\rm\ the intensity pattern, $I_1(\rhovec_1,t)$, that would have illuminated the pinhole detector in the usual lensless ghost  imaging configuration.  Furthermore, we can subtract the time average of this intensity, in our computation, and obtain the equivalent of $\Delta I_1(\rhovec_1,t)$.  To distinguish these computed values from actual light-field quantities, we will denote them $\tilde{I}_1(\rhovec_1,t)$ and $\Delta\tilde{I}_1(\rhovec_1,t)$, respectively.  The time average correlation function,
\begin{eqnarray}
\Delta\tilde{C}(\rhovec_1) &\equiv& \left\langle \int\!d\tau_1\,q\eta_1A_1\Delta\tilde{I}_1(\rhovec_1,t-\tau_1)h_1(\tau_1)\right.\nonumber \\[.08in]
&\times& \left.\int\!d\tau_2\,q\eta_2P_2(t-\tau_2)h_2(\tau_2)\right\rangle_{T_a}
\end{eqnarray}
will then be a background-free ghost image---with spatial resolution $\lambda_0L/D$ over a spatial extent of width $\lambda_0L/d$---akin to what would  be obtained from pseudothermal ghost imaging with $d \approx \rho_0$, $D\approx a_0$, and a DC block applied to the pinhole detector.  Now, because only one photodetector has been employed, it is \em impossible\/\rm\ to interpret this computational ghost image as arising from nonlocal two-photon interference.  
 \begin{figure}[t]
\begin{center}
\includegraphics[width= 3.25in]{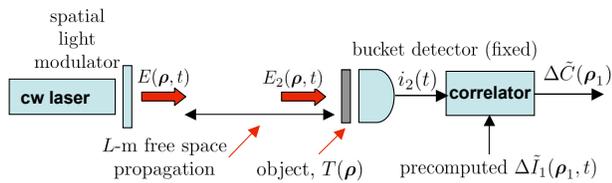}
\end{center}
\vspace*{-.25in}
\caption{(Color online) Computational ghost-imaging setup.} 
\end{figure}

In addition to obviating the need for a high spatial-resolution detector in ghost-image formation, at the expense of the computational burden associated with the free-space propagation calculation for the 
field $E(\rhovec,t)$, computational ghost imaging permits 3D sectioning to be performed.  To see that this is so, we return to the Fig.~1 ghost-imaging configuration with pseudothermal light and inquire about its depth of focus,.  In other words, how badly is the ghost image blurred if the object is at $z=L$ but the pinhole detector is at $z=L + \Delta L$?  This question is easily answered.  Equation~(\ref{fluct}) becomes
\begin{eqnarray}
\lefteqn{\langle\Delta I_1(\rhovec_1,t_1)\Delta I_2(\rhovec_2,t_2)\rangle = }\nonumber \\ &&
|K''(\rhovec_1,\rhovec_2)|^2 |R(t_2-t_1-\Delta L/c)|^2|T(\rhovec_2)|^2,
\label{fluctnew}
\end{eqnarray}
where
\begin{equation}
K''(\rhovec_1,\rhovec_2) \equiv \int\!d\rhovec\,K'(\rhovec,\rhovec_2)
\frac{ik_{0} e^{-i k_{0} (\Delta L+ |\rhovec-\rhovec_1|^2/2\Delta L)}}{2 \pi \Delta L}.
\end{equation}
As a result, the ghost-image term from Eq.~(\ref{Gimage}) takes the same form with $\rho_L$ replaced by $\rho'_L \equiv \rho_L\sqrt{1+ (\Delta L/k_0\rho_L^2)^2}$,
so that the focal region is $|\Delta L| \le k_0\rho_L^2 = 4L^2/k_0a_0^2$.  In the near-field of the pre-diffuser laser beam, i.e., when $k_0a_0^2/4L \gg 1$, the focal region is a very small fraction of the source-to-object path, because $|\Delta L|/L = 4L/k_0a_0^2 \ll 1$ as reported for the experiments in \cite{Meyers}.    This limited depth of focus has the following implications when a range-spread opaque object is imaged in reflection.  The pseudothermal ghost imager can only image one focal region at a time.  However, because the computational ghost imager can precompute $\Delta\tilde{I}_1(\rhovec_1,t)$ for a wide range of propagation distances, the same bucket-detector photocurrent can be correlated with many such $\Delta\tilde{I}_1(\rhovec_1,t)$ to perform 3D sectioning of the object's reflectance.  Of course, this sectioning further increases the computational burden, but this burden can be handled off-line, and, for a given SLM and its associated modulation waveforms, the \em same\/\rm\ precomputed $\Delta\tilde{I}_1(\rhovec_1,t)$ can be used for \em all\/\rm\ ghost images formed using that system.   

In conclusion, we have shown that ghost imaging can be performed with only a bucket (single-pixel) detector by precomputing the intensity fluctuation pattern that would have been seen by the high spatial-resolution detector in lensless ghost imaging.  This computational ghost imager yields background-free images whose resolution and field of view can be controlled by choice of spatial light modulator parameters, and it can be used to perform 3D sectioning.  Finally, the computational ghost imager underscores the classical nature of ghost-image formation.  In particular, as only one light beam and one detector are employed, it is \em impossible\/\rm\ to consider the computational ghost image as arising from nonlocal two-photon interference, as has been argued in \cite{Scarcelli,Shih} for the pseudothermal case.  

This work was supported by the U. S. Army Research Office MURI Grant W911NF-05-1-0197, the DARPA Quantum Sensors Program, and the W. M. Keck Foundation for Extreme Quantum Information Theory.


\begin{thebibliography}{10}

\bibitem{Pittman} T. B. Pittman {\em et al.}, Phys. Rev. A {\bf 52,} R3429 (1995).

\bibitem{Valencia} A. Valencia {\em et al.}, Phys. Rev. Lett. {\bf 94,} 063601 (2005).

\bibitem{Ferri} F. Ferri {\em et al.}, Phys. Rev. Lett. {\bf 94,} 183602 (2005).

\bibitem{Gatti:three} A. Gatti {\em et al.}, Phys. Rev. A {\bf 70,} 013802 (2004).

\bibitem{Gatti} A. Gatti {\em et al.}, Phys. Rev. Lett. {\bf 93,} 093602 (2004).

\bibitem{Mandel} L. Mandel and E. Wolf, \em Optical Coherence and Quantum Optics\/\rm\ (Cambridge Univ., Cambridge, 1995), chapters.~4,~9,~12.

\bibitem{DAngelo} M. D'Angelo {\it et al.}, Phys. Rev. A {\bf 72,} 013810 (2005).

\bibitem{Cai} Y. Cai and S.-Y. Zhu, Opt. Lett. {\bf 29,} 2716 (2004).

\bibitem{Cai2} Y. Cai and S.-Y. Zhu, Phys. Rev. E {\bf 71,} 056607 (2005).

\bibitem{Scarcelli} G. Scarcelli, V. Berardi and Y. Shih, Phys. Rev. Lett. {\bf 96,} 063602 (2006).

\bibitem{ErkmenShapiro}B. I. Erkmen and J. H. Shapiro, Phys. Rev. A {\bf 77,} 043809 (2008).

\bibitem{Meyers}R. Meyers, K. S. Deacon, and Y. Shih, Phys. Rev. A {\bf 77,} 041801(R) (2008).

\bibitem{Shih}Y. Shih, arXiv:0805.1166 [quant-ph].  

\bibitem{Goodman}J. W. Goodman, \em Introduction to Fourier Optics\/\rm\ (McGraw Hill, New York, 1968), chapter~4.

\bibitem{footnote1}Alternatively, we could have illuminated a rough-surfaced opaque object and used the bucket detector to collect a fraction of the light that is diffusely reflected therefrom, cf.\ the experiment reported in \cite{Meyers}.  Because the physics we are after is not affected by whether we perform ghost imaging in transmission or reflection, we have chosen to adhere to the transmissive configuration.  

\bibitem{coherence_separability} Although the sources in ghost imaging experiments may not be cross-spectrally pure, this assumption simplifies the analytical treatment without compromising the fundamental physics of image formation.

\bibitem{Wozencraft}J. M. Wozencraft and I. M. Jacobs, \em Principles of Communication Engineering\/\rm\ (Wiley, New York, 1965), chapter~3.  

\bibitem{MeyersRes}This behavior explains why better ghost-imaging spatial resolution was reported in \cite{Meyers} when the size of the pseudothermal source was increased.  

\bibitem{background}We show in \cite{ErkmenShapiro} that the contrast of this ghost image is $\sim$$1/N$, for an object transparency satisfying $|T(\rhovec)|^2 = 0$ or 1, where $N$ is the number of resolution cells on the object.  

\bibitem{telegraph}These statistics prevail if $\{e^{i\phi_{nm}(t)}\}$ is a set of iid random telegraph waves, see, e.g., A. Papoulis, \em Probability, Random Variables, and Stochastic Processes, Third Ed.\/\rm\ (McGraw Hill, New York, 1991), chapter~10.

\bibitem{reflect}It is also possible to perform computational ghost imaging in reflectance, cf. \cite{Meyers} for the pseudothermal case.
\end{thebibliography}
\end{document}